\documentclass[onecolumn]{aa} 
\usepackage{graphicx,natbib}
\usepackage{txfonts}
\usepackage{dcolumn}
\usepackage{bm}

\newcommand{\vect}{\vec}

\newcommand{\dpth}{\partial_{\theta}}
\newcommand{\dpph}{\partial_{\phi}}

\newcommand{\dpht}{\partial_{\hat t}}
\newcommand{\dphr}{\partial_{\hat r}}
\newcommand{\dphth}{\partial_{\h \theta}}
\newcommand{\dphph}{\partial_{\h \phi}}

\newcommand{\dphrr}{\partial^2_{\hat r}}
\newcommand{\dphtt}{\partial^2_{\hat t}}
\newcommand{\diver}{{\rm div~}}
\newcommand{\rot}{{\rm curl~}}

\newcommand{\h}{\hat}
\newcommand{\de}{\delta}

\begin{document}

\title{Non-stationary Rayleigh-Taylor instability in supernova ejecta}
\author{X. Ribeyre \inst{1} \and L. Hallo \inst{1} \and V.T. Tikhonchuk \inst{1} \and S. Bouquet \inst{2} \and  J. Sanz \inst{3} }
\institute{Centre Lasers Intenses et Applications, UMR 5107 CNRS -
Universit\'e Bordeaux 1 - CEA, Universit\'e Bordeaux 1, 351, Cours de la Lib\'eration, 33405 Talence Cedex, France
\and Commissariat \`a l'Energie Atomique, DIF/D\'epartement de Physique 
Th\'eorique et Appliqué\'ee, 91680, Bruy\`eres le Ch\^atel, France
\and E.T.S.I., Aeron\'auticos, Universidad Polit\'ecnica de Madrid, Madrid 28040, Spain}

\abstract{
The Rayleigh-Taylor instability plays an important role in the
dynamics of several astronomical objects, in particular, in 
supernovae (SN) evolution. In this paper we develop an analytical
approach to study the stability analysis of spherical expansion of the
SN ejecta by using a special transformation in the co-moving
coordinate frame. We first study a non-stationary spherical
expansion of a gas shell under the pressure of a central source.
Then we analyze its stability with respect to a no radial, non spherically symmetric perturbation of the of the shell. We consider the case where the polytropic constant of the SN shell is $\gamma=5/3$ and we examine the evolution of 
a arbitrary shell perturbation.  The dispersion relation is derived.
The growth rate of the perturbation is found and its temporal and
spatial evolution is discussed. The stability domain depends on
the ejecta shell thickness, its acceleration, and the perturbation
wavelength.}

\maketitle
\section{Introduction}
The Rayleigh-Taylor Instability (RTI) is a common phenomena in 
supernovae evolution. For a type II supernova (SN), the
neutrino-driven RTI is predicted during the core collapse. It is
excited at the discontinuity surface between the stalled shock
wave and the neutrino sphere~\citep{Janka96}. In supernovae
interiors, for example for the SN 1987A~\citep{Fryxell91}, the RTI
is thought to be responsible for the mixing of the heavy elements (Ni,
Si) with the light ones (He, H). This mixing process allows one
to obtain a better interpretation of the light curve of the SN
1987A.

In this paper we consider the type II SN explosion. When the thermonuclear 
reactions cannot supply anymore the energy required to maintain 
the pressure equilibrium in the star, its
core collapses and produces a neutron star at its center. If the
rotation axis and the magnetic field lines are not
parallel inside the neutron star, the former becomes a pulsar
 which gradually converts the spinning energy into a flux of electromagnetic waves and high energy particles. 
This pulsar wind blows up the inner layers of the
supernova and creates a dense shell expanding in the circumstellar medium
 -- the SN remnant (SNR). This structure of a shell expanding under the pressure of a central pulsar is called a plerion~\citep{Weiler80}.

This object experiences several hydrodynamic
instabilities, especially, the inner shell surface is RT unstable during the acceleration phase. It is suggested that the RTI at the inner ejecta surface is responsible for the shell fragmentation and for the filamentary structure in the Crab nebula~\citep{Hester96}. For type Ia and type II SNe, the RTI arises also at the outer surface of an old SNR when the ejected shell is decelerated by the interstellar medium~\citep{Velazquez98,Chevalier92,Herant91}. The plerion stability analysis is complicated for the pulsar pressure is non stationary, the density distribution in the shell is not uniform and, in addition, one has to take into account the velocity of the shell since it is in rapid expansion.

Although the RTI has been studied for a long time since the
pioneering work of Lord Rayleigh~\citep{Rayleigh83} and Sir G. Taylor~\citep{Taylor50}, only a limited number of exact analytical solutions is known. However, the application of the RTI to the SN explosion, has been considered earlier by Bernstein and Book~\citep{Bernstein78}, Reynolds and Chevalier~\citep{Reynolds84} and Blondin et al.~\citep{Blondin01}. Similar studies about the stability of target implosions have also been performed in the field of inertial confinement fusion~\citep{Hattori86, Han91, Goncharov00} (ICF). In
these both domains the hydrodynamic instabilities develop 
in spherical geometry. Although the spatial and 
temporal scales as well as signs of the shell velocity and 
acceleration are different, the qualitative results of both series 
of publications are comparable.

However, the stability analysis is not completed yet.  The study by
Hattori et al.~\citep{Hattori86} is limited to the stagnation
period of the shell evolution and it assumes a uniform density
profile. Bernstein and Book~\citep{Bernstein78} carried out the
study of the RTI for a shell in expansion, but they considered only
an asymptotic limit of the RTI growth rate. Blondin et al.~\citep{Blondin01} considered only the instabilities that occur in
the late phase of the SN evolution, when the ejecta shell is
decelerated outside the pulsar nebula. The present paper provides another aspect of the study, i.e., the early stage of the SNR evolution when the shell is accelerated (and its thickness increases - see later) by the pulsar wind.
On the other hand, this approach works also for the description in the compression of a collapsing shell. 

In this paper, Section~\ref{Sec:2} provides the basic equations 
in the co-moving frame. Trough this procedure, we transform the non-stationary evolution of the shell into a steady motion. We perform the stability analysis for the case of an ideal mono-atomic gas. An analytical  
solution for the unperturbed spherically-symmetric flow is given  
in Section~\ref{Sec:3}. We present the geometry and discuss about the physical parameters relevant to the conditions of a pulsar nebula-supernova 
remnant interaction. The stability analysis is performed in the 
co-moving coordinate frame and the dispersion relation is obtained
in Section~\ref{Sec:4}. Section~\ref{Sec:5} is devoted to the 
numerical study of the dispersion relation and the nature of the unstable solution is studied in details in section~\ref{Sec:6}. The temporal evolution of the shell outer and inner surfaces are analyzed in Section~\ref{Sec:7}. The conclusions are given in Section~\ref{Sec:8}. 
 
\section{Zooming coordinates}\label{Sec:2}
Let us consider a spatio-temporal evolution of the shell
expanding under the central pulsar pressure. Our model equations consist in 
the continuity and Euler equations for the gas density $\rho$ and the velocity 
$\vec v$: 
\begin{eqnarray}
&&\partial_t \rho+\vec \nabla_{\vec r}\cdot(\rho \vec v) = 0, \label{equ:1}\\
&&\partial_t \vec v+ (\vec v\cdot\vec\nabla_{\vec r})\vec v = -\frac{1}{\rho} \vec\nabla_{\vec r} \,p ,\label{equ:2}
\end{eqnarray}
for an ideal gas with a polytrope equation of state $p=K\, \rho^{\gamma}$ and where the subscripts ``$t$'' and ``$\vec r$'' stand for the partial derivative. The differential operators in the spherical coordinates $(r,\theta,\phi)$ read:
\begin{eqnarray*}
&&\vec \nabla_{\vec r}\cdot(\rho \vec v) =\frac1{r^2} \, \partial_r (r^2\rho v_r) + \frac1{r\,\sin\theta} \, \partial_\theta (\sin\theta \rho v_\theta)  \\ 
&& + \frac1{r\,\sin\theta} \, \partial_\phi (\rho v_\phi),  \\
&&\vec \nabla_{\vec r}\, p =\{ \partial_r p,\, \frac1{r} \, \partial_\theta p,\, \frac1{r\,\sin\theta} \, \partial_\phi p\},\\
&&(\vec v\cdot\vec\nabla_{\vec r}) v_r=v_r\partial_r v_r+ \frac{ v_\theta}{r} \, \partial_\theta  v_r + \frac{ v_\phi}{r\,\sin\theta} \, \partial_\phi  v_r - \frac1r\,(v_\theta^2+ v_\phi^2),\\
&&(\vec v\cdot\vec\nabla_{\vec r}) v_\theta=v_r\partial_r v_\theta + \frac{ v_\theta}{r} \, \partial_\theta  v_\theta + \frac{ v_\phi}{r\,\sin\theta} \, \partial_\phi  v_\theta \\
&& + \frac1r\,(v_r v_\theta- \cot\theta\, v_\phi^2), \\
&& (\vec v\cdot\vec\nabla_{\vec r}) v_\phi=v_r\partial_r v_\phi + \frac{ v_\theta}{r} \, \partial_\theta  v_\phi + \frac{ v_\phi}{r\,\sin\theta} \, \partial_\phi  v_\phi  \\
&& +\frac1r\,(v_r v_\phi+ \cot\theta\, v_\theta v_\phi),
\end{eqnarray*}
where $v_r, v_\theta$ and $v_\phi$ are respectively the radial, azimuthal and tangential component of the velocity $\vec v$.
Even for a radially symmetric flow this system of equations is  
difficult to solve because of the non-stationary character of the evolution.

Eqs. (\ref{equ:1}) and (\ref{equ:2}) are solved by using a time-dependent transformation initially introduced by Munier and Feix~\citep{Munier83} and, thereafter by Bouquet et al.~\citep{Bouquet85, Blottiau88} and nowadays referred in astrophysics as 'zooming coordinates'~\citep{Hanawa00,Hennebelle01,Shadmehri01}. It is a specific coordinate transformation from the laboratory
space to the co-moving, radially expanding frame. This new space 
is labelled with a hat symbol and it is defined by the following 
relations:
\begin{eqnarray}
&&r=C(t)\,\h r,\quad \theta=\h \theta,\quad \phi=\h \phi, \quad dt=A^2(t)\, d\h t \label{equ:3},\\
&&\rho=D(t) \, \h \rho, \quad p=B(t) \, \h p \label{equ:4},
\end{eqnarray}
where $A,\,B,\,C,\,D$ are the time dependent scaling functions; $\h \rho$ and $\h p$ are the 
density and pressure in the co-moving frame depending on $\h r,\h\theta,\h\phi$ and $\h t$. The velocity $\h{\vec v}$ in the new frame is given by the classical definition: 
\begin{eqnarray*}
&&\vec{\h v} =  \frac{d\vec{\h r}}{d\hat t}.
\end{eqnarray*}
According to Eq. (\ref{equ:3}), the velocity transformation reads:
\begin{eqnarray}
&&\vec v=\frac{C}{A^2}\hat{\vec v} + \dot C \hat{\vec r}, \label{equ:5}
\end{eqnarray}
where the dot stands for the time derivative. These two reference frames coincide at $t=\hat t=0$, that is: $A(0)=B(0)=C(0)=D(0)=1$. In the new frame the equation of state reads $\hat p=(D^\gamma/B) K \hat \rho^{\gamma}$ and Eqs. (\ref{equ:1}) and (\ref{equ:2}) become:
\begin{eqnarray}
&&\partial_{\hat t} \hat\rho+\vec \nabla_{\hat{\vec r}}\cdot(\hat\rho \hat{\vec v}) = - \Omega_1\hat\rho, \label{equ:6}\\
&&\partial_{\hat t} \hat{\vec v}+ (\hat{\vec v}\cdot\vec\nabla_{\hat{\vec r}})\hat{\vec v} = -\frac{N}{\hat\rho} \vec\nabla_{\hat{\vec r}}\, \hat p -\Omega_2 \h{\vec v} -\frac1{\tau^2}\, \hat{\vec r}, \label{equ:7}
\end{eqnarray}
where $\Omega_1,\,N,\,\Omega_2$ and $\tau$ are the source terms arose due to the non-inertial nature of the new reference frame. We have:
\begin{eqnarray*}
&&\Omega_1=A^2\left(3\, \frac{\dot C}{C}+\frac{\dot{D}}{D}\right),\, N=\frac{BA^4}{C^2D},\\
&&\Omega_2= 2A^2\left(\frac{\dot{C}}{C}-\frac{\dot{A}}{A}\right) , \quad \frac{1}{\tau^2}=\frac{\ddot{C}A^4}{C}, 
\end{eqnarray*}
where $\Omega_i (i=1,2)$ has the dimension of a frequency, $\tau$ is a time and $N$ is a dimensionless number.
The scaling functions $A,\,B,\,C,\,D$ can be found from invariance
considerations and conservation laws.
 
First of all, we are going to require that the EOS in the new frame remains invariant compared to the EOS in the initial frame. In other words, the fluid keep the same thermodynamical properties in the new space as in the physical one. The condition $\h p=K\h \rho^\gamma$ implies that $D^\gamma/B=1$ and, therefore, $B=D^\gamma$. This is the first constraint for the scaling function. 

Second, requesting the same property for the continuity equation, i.e., the amount of matter is preserved from one space to the other, one deduces $\Omega_1=0$ and we obtain $D=(1/C)^3$. The single equation left is the equation of motion. In order to keep invariant the pressure gradient force which originates from fundamental physics laws in the initial space, we take $N=1$. Combining the former three relations for the scaling functions, one has \citep{Ribeyre05}: 
\begin{eqnarray}
&A=C^{(3\gamma-1)/4}, \quad B=C^{-3\gamma}, \quad D=C^{-3}, \label{equ:8}
\end{eqnarray}
At this stage, we have expressed the three scaling  functions
$A,\,B$, and $D$ in terms of the single one, $C$, which is the function
governing the scaling between the radial coordinates $r$ and $\h
r$. However we have to satisfy two more conditions: the
coefficients in the friction term and in the radial forces in Eq.
(\ref{equ:7}) become respectively:
\begin{eqnarray}
&&(3\,\gamma-5)\,\dot{C}\, C^{3(\gamma - 1)/2}=2\,\Omega_2, \label{equ:9} \\
&&\ddot{C}\,C^{3\gamma-2} = \tau^{-2}. \label{equ:10}
\end{eqnarray}
In order to have a stationary flow in the co-moving  frame, both
parameters $\tau$ and $\Omega_2$ should be constant. An inspection of
Eqs.~(\ref{equ:9}) and (\ref{equ:10}) shows that the only way to
satisfy these two constraints is to set $\Omega_2 = 0$ by choosing the
polytropic constant $\gamma= 5/3$  which corresponds to
a mono-atomic ideal gas. Then, the friction force (proportional to $\h{\vec v}$) vanishes  in Eq.~(\ref{equ:6}). Finally, the scaling function $C(t)$ is obtained from the resolution of the differential equation~(\ref{equ:10}) and $A(t)$, $B(t)$ and $D(t)$ are derived from Eq.~(\ref{equ:8}). This is the simplest case, which allows an analytical study of the instability development for a non-stationary regime in the space ($r, \theta, \phi, t$).

For another value of $\gamma$ ($\gamma \neq 5/3$), the scale function $C$ is still defined by Eq. (\ref{equ:10}), but then the function $\Omega_2(t)$ is neither zero nor constant and it accounts for an effective friction force in the co-moving frame. However,  this approach is not very fruitful since a time-dependant coefficient appears in the rescaled equations and the problem becomes more complex in the ($\h r, \h t$)-space than the initial one.

This result is clearly in apposition with our procedure which consists in making more simple a initial complex problem.

\section{Unperturbed flow solution}\label{Sec:3}

With this choice of $\gamma$ all coefficients in (\ref{equ:6}) and (\ref{equ:7}) are constants and one may look for a static, radially-symmetric solution for these equations in the co-moving frame. It can be found by setting $\h{\vec{v}}(\h r,\h t)\equiv \h{\vec v}_0=0$, $\h\rho(\hat r, \h t)\equiv \h\rho_0(\hat r)$, and $\h p(\hat r,\h t)\equiv \h p_0(\hat r)$.  
Then the solution reads:
\begin{eqnarray}
&&\h\rho_0(\h r)=\h\rho_0(0)\,\left(1-\frac{\h r^2}{\h r_1^2}\right)^{1/(\gamma-1)} \quad \mbox{and} \nonumber \\
&&\h p_0(\h r)=K \h\rho_0^\gamma(0)\,\left(1-\frac{\h r^2}{\h r_1^2}\right)^{\gamma/(\gamma-1)}, \label{equ:11}
\end{eqnarray}
where the central density  $\h\rho_0(0)=[\h r_1^2\,(\gamma-1)/
2K\gamma\tau^2]^{1/ (\gamma-1)}$. This solution is spatially
bounded for $\tau^2>0$ in a sphere of radius $\h r_1$. To
return to the laboratory frame one must find the scaling
function $C$ from Eq. (\ref{equ:10}).

\begin{figure}
\centerline{\includegraphics[width=8cm]{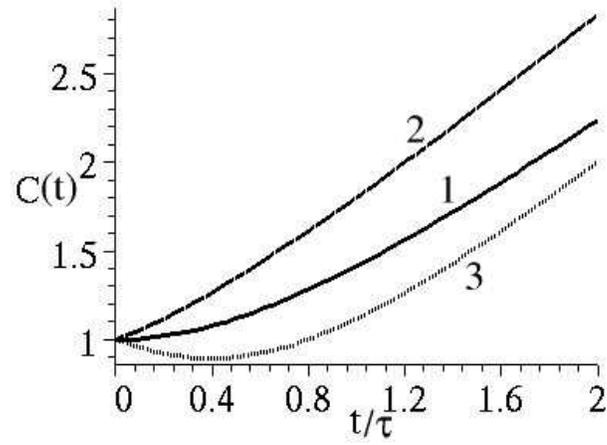}}
\caption{Scaling function $C(t)$ in three cases:  without initial
velocity, $\beta = 0$ (1), with a positive initial velocity,
$\beta= 0.5$ (2) and with a negative initial velocity, $\beta =
-0.5$ (3).\label{fig1}}
\end{figure}

In the case of a mono-atomic  gas where $\gamma = 5/3$,  the
solution to Eq. (\ref{equ:10}) is fully analytic and it is of most interest for a wide class of astronomical objects~\citep{Bernstein78, VanderSwaluw04}. The scaling function $C$ reads:
\begin{eqnarray}
&&C(t) = \sqrt{\left(1+\frac{\beta t}{\tau}\right)^2+\frac{t^2}{\tau^2}}. \label{equ:12}
\end{eqnarray}
where $C(0)=1$ and the constant $\beta \equiv \tau \dot C(0)$
characterizes the shell initial velocity $v_0(r)=\beta r /\tau$ is the initial physical ($\vec r,t$) space.
The function $C(t)$ is plotted in Fig.~\ref{fig1} for three
initial velocities. In all three cases $C$ approaches the
ballistic motion [$C(t) \propto t$] for $t\gg\tau$, but for
$t<\tau$ the behaviors are quite different. If the initial velocity is
positive or zero, $C(t)$ is monotonic and the flow is always in
expansion. If the initial velocity is negative, $\beta < 0$, there
is a minimum radius, which corresponds to a stagnation at the time
$t_r=\tau\,|\beta|/(\beta^2+1)$, followed by an expansion regime.

\begin{figure}
\centerline{\includegraphics[width=8cm]{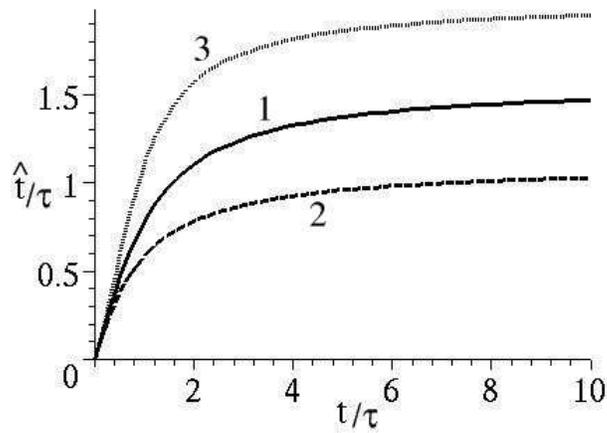}}
\caption{Relation between the time $\h t$ in the co-moving frame
and the time $t$ in the laboratory frame, for $\beta = 0$ (1), $0.5$ (2), and $-0.5$ (3).\label{fig2}}
\end{figure}

Once the function $C(t)$  is found, the integration of the fourth
equation in (\ref{equ:3}) provides a relation between the two times $t$ and $\h t$:
\begin{eqnarray}
&&\h t = \tau g(t) \nonumber \\
&&\mbox{where}\quad g(t)= \arctan [\beta+(\beta^2 +1)\,t/\tau]-\arctan\beta. \label{equ:13}
\end{eqnarray}
Figure \ref{fig2} shows the relation between $\h t$ and $t$ for three values of $\beta$. Although $t$ varies from zero to infinity in the laboratory frame, $\h t$ is bounded and it varies in the range $[0,\h t_{max}[$ where the upper value $\h t_{max}$ depends on $\beta$ and is given by $\h t_{max}= \tau (\pi/2-\arctan\beta)$.

As the function $C$ is known, one obtains the other scaling functions $A(t)$, $B(t)$ and $D(t)$ from  Eqs.~(\ref{equ:8}). As a consequence, the non-stationary solution in the laboratory frame can be derived and we get:
\begin{eqnarray}
&&\rho(r,t)=\h\rho_0(0)\,C^{-3} \,\left(1-\frac{r^2}{C(t)^2\,\h r_1^2}\right)^{3/2} \label{equ:14},\\
&&p(r,t)=K\,\h\rho_0^{5/3}(0)\,C^{-5}\,\left(1-\frac{r^2}{C(t)^2\h r_1^2}\right)^{5/2} \,  \label{equ:15},\\
&&v_r(r,t)=\frac{r}{C^2\tau}\,\left(\beta + (\beta^2+1)\frac{t}{\tau}\right). \label{equ:16}
\end{eqnarray}
The solution describes a bubble expansion, with density and pressure decreasing with time while the radius increases. The velocity increases linearly within the bubble. The flow is compressible and its divergence,
\begin{eqnarray}
&&\diver  \vect{V}= \frac{3\dot C}{C}=\frac{3}{\tau}\frac{(\beta+t/\tau)}{[(1+\beta t/\tau)^2+t^2/\tau^2]} \label{equ:15bis},
\end{eqnarray}
decreases with time as $3/t$ when $t \rightarrow \infty$, i.e., the volume of the flow grows with time. This solution could
be found in a different way by imposing initially the radial
velocity profile~\citep{London86} : $v_r \propto r$. However, the
method we used here is more efficient to perform the perturbation
analysis presented below.

\begin{figure}
\centerline{\includegraphics[width=11cm]{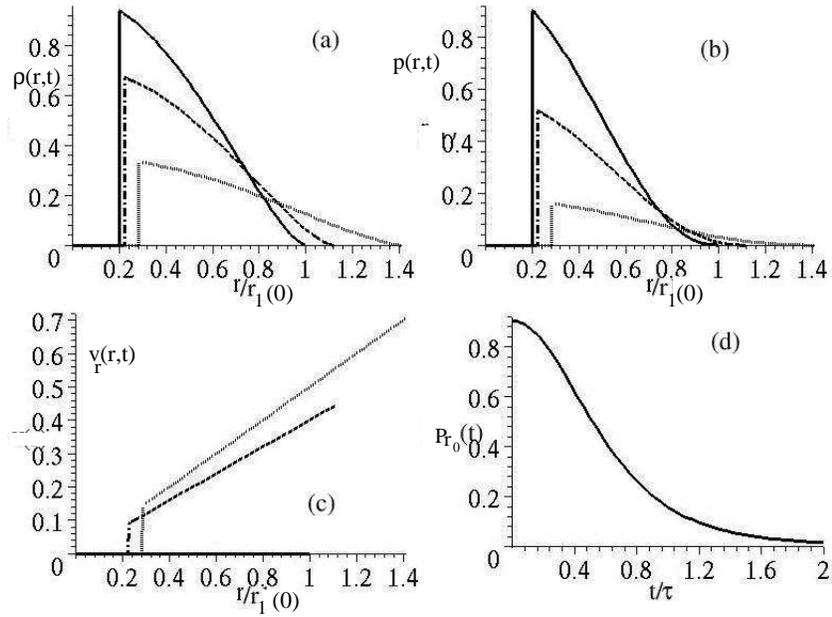}}
\caption{Density (a), pressure (b) and velocity (c) profiles
versus the radius in the laboratory frame at $t=0$ (solid
line), $t=0.5\,\tau$ (dotted-line), and $t=\tau$ (dashed-line), (d) 
plot of the pulsar pressure $p_{r_0}$ at the inner surface of the 
shell versus time. The initial parameters are $\beta = 0$ and $\h r_0=0.2\,\h r_1$. The velocity is normalized by $\h r_1/\tau$, the density -- by the
central density $\h \rho_0(0)$ and the pressure -- by $K \h \rho_0(0)^{5/3}$. \label{fig3}}
\end{figure}

From the above solution one can construct a model for a supernova 
remnant blown-up by a pulsar wind. For that we remove the inner part of the
bubble within the  sphere of radius $\h r_0 < \h r_1$ in the
co-moving frame. The pressure of the removed fluid is replaced by
a radiative pressure, on the inner surface of the shell, 
attributed to the pulsar wind. The solution
corresponding, therefore, to an expanding shell in the laboratory frame is
shown in Fig.~\ref{fig3}. The inner (resp. outer) radius is given
by $r_0(t)=C(t)\,\h r_0$ [resp. $r_1(t)=C(t)\,\h r_1$].

The density, pressure and velocity profiles in the shell follow
directly from Eqs. (\ref{equ:14}) - (\ref{equ:16}). They are given
in Fig.~\ref{fig3}.  The shell thickness $L$ increases as 
$L = C\,L_0$ where $L_0=\h r_1- \h r_0$ is the initial shell 
thickness. The pressure law at the inner interface of the shell 
describes the pressure produced by the pulsar wind:
\begin{eqnarray}
&&p_{r_0}(t)=K\,\h\rho_0^{5/3}(0)\,C^{-5}\,\left(1-\frac{r_0^2(0)}{r_1^2(0)}\right)^{5/2}. \,  \label{equ:17}
\end{eqnarray}
Figure \ref{fig3}(d) shows the time evolution of the pressure at 
the inner surface of the shell. In particular, for $t\gg\tau$ the 
pressure decreases as $p_{r_0}\propto t^{-5}$. This temporal 
dependence of the pressure can be related to the pulsar 
luminosity: $L(t) \propto - d(p_{r_0} V)/dt$, where $V$ is the 
volume of the inner cavity bounded by the shell. Since the volume increases as $C^3$, the derivative of the product $p_{r_0} V$ varies according to $\dot C/ C^{-3}$. As a result, asymptotically for  $t\gg \tau$, one finds that $L(t) \propto t^{-3}$. This dependence 
is in good agreement with the classical model of pulsar luminosity. Indeed,
according to Blondin et al.~\citep{Blondin01}, for $t\gg\tau$ the 
pulsar luminosity decreases as $t^{-\mu}$ where $2<\mu <3$ and $\tau$ 
is the characteristic life time of the pulsar~\citep{Camilo00}. 
Consequently, the  pressure law which follows from our analytical 
solution (\ref{equ:17}) is  relevant for a class of pulsars  experiencing a
slowing down of their rotation velocity.

In addition, the radial distribution of the flow velocity in our model is also in agreement with observations. For instance, the motion and structure of the filamentary envelope of the Crab nebula has been studied by Trimble~\citep{Trimble68, Cadez04}. The motion of filaments is largely radial and the velocity of each filament is approximately proportional to its distance from the expansion center, $v \propto r$.  On the other hand, these observations indicate that the filaments have been accelerated. It turns out that this behavior can be probably fitted with our analytical model by an appropriate adjustment of the parameters $\beta$ and $\tau$.

\section{Study of the shell stability}\label{Sec:4}

\begin{figure}
\centerline{\includegraphics[width=8cm]{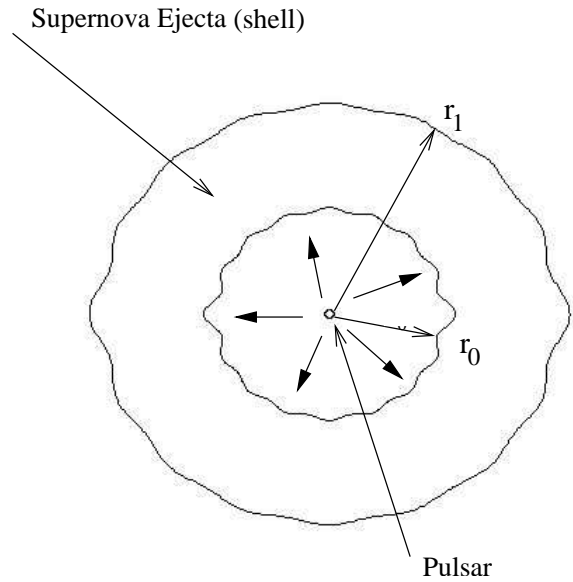}}
\caption{Sketch of the shell explosion under the pulsar
pressure. The inner, $r_0=\h r_0 \, C(t)$, and outer, $r_1=\h
r_1\, C(t)$, radii of the shell vary with time. The inner surface
is pushed by the pulsar wind and it is RT unstable.\label{fig4}}
\end{figure}

In this section we study the linear stability of the expanding shell against Rayleigh-Taylor (RT) type perturbations, assuming that the shell is accelerated by the pulsar wind pressure. As a matter of fact, this configuration is RT unstable for the light fluid (radiation) which pushes the heavy fluid (shell of the supernova remnant). As it is shown in Fig.~\ref{fig4}, small perturbations at the inner surface might grow during the acceleration phase and eventually break down the shell.

The stability analysis of the shell in the laboratory frame is
complicated since the background flow is non-stationary and the
density and pressure vary with time and space. This analysis 
is much more simple in the co-moving frame since the unperturbed shell there is at rest. We define any perturbed physical quantity $\h q(\h r,\h \theta, \h \phi, \h t)= \h q_0(\h r) + \de\h q(\h r,\theta, \phi, 
\h t)$ where $\de\h q(\h r,\theta, \phi, \h t)$ is the perturbation in the ($\vec{\h r},\h t$)-space, and the perturbed position of the inner boundary is $\delta \h r = \h \eta(\theta, \phi, \h t)$.  Since $\Omega_1=0$ and assuming $\gamma =5/3$, i.e., $\Omega_2=0$, one then obtains the following set of linearized hydrodynamic equations [Eqs. (\ref{equ:6})-(\ref{equ:7})]:
\begin{eqnarray}
&& \dpht \de \h \rho + \frac{1}{\h r^2} \dphr (\h r ^2 \h \rho_0 \de \h v_r) + \frac{1}{\h r \sin \theta} \, \dphth (\sin \theta \h \rho_0 \de \h v_\theta) \nonumber \\
&&+ \frac{1}{\h r \sin \theta} \, \dphph (\h \rho_0 \de \h v_\phi) = 0, \label{equ:18}\\
&& \h \rho_0 \, \dpht \de \h v_r = -\dphr \de \h p - \tau^{-2} \h r \, \de \h \rho, \label{equ:19}\\
&& \h r\,\h \rho_0 \, \dpht \de \h v_\theta = - \dpth \de \h p, \label{equ:20}\\
&& \h r \sin \theta\, \h \rho_0 \, \dpht \de \h v_\phi = - \dpph \de \h p, \label{equ:21}
\end{eqnarray}
where $ \de \h p=C^2_{s0}(\h r) \de \h \rho$ and $C^2_{s0}(\h r)=(5/3) K\h 
\rho^{2/3}_0(\h r)\equiv (\h r_1^2- \h r^2)/3\tau^2$ is the square of the sound speed. 

One needs also to take into account the boundary conditions at the interfaces $\h r_0$ and $\h r_1$. Let us assume that each of interfaces undergoes the perturbation $\h \eta_i$ ($i=0,1$) . Then one should request the continuity of the pressure and the radial velocity at each interface, $\h r = \h r_i+ \h \eta_i$, that is,
\begin{eqnarray}\label{equ:22}
&&\dpht \h \eta_i = \delta \hat v_r(\h r_i), \quad
\partial_{\h r} \h p_0 (\h r_i)\,\h \eta_i + \delta \h p(\h r_i) = 0, \quad i = 0,1.
\end{eqnarray}

Introducing the relative density perturbation (density contrast) $\h \epsilon = \de \h \rho / \h \rho_0$, one can reduce the system of Eqs. (\ref{equ:18}), (\ref{equ:19}), (\ref{equ:20}), and (\ref{equ:21}) to a single second order partial differential equation (PDE):
\begin{eqnarray}
&&\dphtt \h \epsilon + \frac{2}{\tau^2} \h \epsilon  = C_{s0}^2 \dphrr \h \epsilon +\left(\frac{2C_{s0}^2}{\h r}-\frac{7\h r}{3\tau^2}\right)\dphr \h \epsilon +\frac{2\h r^2}{3\tau^4 C_{s0}^2} \h \epsilon+ \nonumber \\
&&\frac{C_{s0}^2}{\h r^2\sin^2 \theta}\,[\sin \theta\, \dpth (\sin \theta \, \dpth \h \epsilon) + \partial^2_\phi \h \epsilon ].\label{equ:24}
\end{eqnarray}
In this PDE, if we keep only the first term of the LHS and the first term of the RHS, as well, one obtain the very simple PDE, $\dphtt \h \epsilon = C_{s0}^2 \dphrr \h \epsilon $, which just corresponds to the sound wave equation, as expected. In Eq.(\ref{equ:24}), much more terms arise and they account for gravity dynamical effect existing in the initial ($\vec r,t$)-space.

From the equations of motion (\ref{equ:19})-(\ref{equ:21}) and from the relation between $\de \h p$ and $\de \h \rho $,  the components of the flow velocity can be expressed in terms of the density contrast:
\begin{eqnarray}
&&\dpht \delta \h v_r =-C_{s0}^2\,\dphr \h \epsilon + \frac{2\h r}{3\tau^2} \h \epsilon,  \label{equ:25}\\
&&\h r \dpht \delta \h v_\theta =-C_{s0}^2\dpth \h \epsilon,  \label{equ:26}\\
&&\h r \sin\theta \dpht \delta \h v_\phi =-C_{s0}^2 \dpph \h \epsilon.  \label{equ:27}
\end{eqnarray}
The angular dependence of the perturbed quantities can be written in terms of spherical harmonics $Y_{lm} (\theta, \phi)$:
\begin{eqnarray}
&&\h \epsilon (\h r, \theta, \phi, \h t) = \h \zeta (\h r, \h t)\, Y_{lm}(\theta,\phi), \label{equ:28} \\
&&\h \eta_{i} (\theta, \phi, \h t) = \h \kappa_{i} (\h t)\, Y_{lm}(\theta,\phi),\, i=0,1\,, \label{equ:29}
\end{eqnarray}
where $\h \zeta (\h r, \h t)$ and $\h \kappa_i (\h t)$; $i=0,1$; are three functions to be determined later on, and where $l$ and $m$ are two integers with $l \geq 0$ and $m \in [-l,l]$. Then Eq.~(\ref{equ:24}) reduces to the equation for the radial function $\h \zeta$: 
\begin{eqnarray}
&&\dphtt \h \zeta + \frac{2}{\tau^2} \h \zeta = \nonumber \\
&&C_{s0}^2 \left [\dphrr \h \zeta +\left(\frac{2}{\h r}-\frac{7\h r}{3C_{s0}^2\tau^2}\right)\dphr \h \zeta +\left(\frac{2\h r^2}{3\tau^4 C_{s0}^4}-  \frac{l(l+1)}{\h r^2} \right) \h \zeta \right]. \label{equ:30}
\end{eqnarray}
This is a linear PDE with time-independent
coefficients. One is therefore allowed to consider an exponential
time dependence for solutions. Denoting $\omega$ the eigenvalue
and $S(\h r)$ the eigenfunction, we represent $\h \zeta(\h r, \h t) 
= (\h a_0/\h r_1) S(\h r) \exp(\omega \h t/\tau)$ and $\h \kappa_i
(\h t) = \h a_i \exp(\omega \h t/\tau)$ ($i=0,1$) where $\h a_i$ is 
the amplitude of the initial perturbation at the inner (i=0) - outer (i=1) 
surface, and we assume that $\h a_i \ll \h r_i$. Then Eq. 
(\ref{equ:30}) can be written in a dimensionless ordinary differential equation (ODE):
\begin{eqnarray}
&&\frac{d^2 S}{dR^2}+\left(\frac{2}{R}-\frac{7R}{1-R^2}\right) \frac{dS}{dR} \nonumber \\
&&+\left(\frac{6(2R^2-1)}{(1-R^2)^2}-\frac{l(l+1)}{R^2}-\frac{3\omega^2}{1-R^2}\right)\,S = 0,\label{equ:31}
\end{eqnarray}
where $R=\h r/ \h r_1$ varies from $R_0=\h r_0/\h r_1$ to $R_1=1$. This 
equation can be reduced to a canonical form with the transformation $S(R)=R^\mu\,F(R^2)/(1-R^2)$:
\begin{eqnarray}
&&x(x-1)\frac{d^2 F}{dx^2}+\left[x(\mu+3)-\left(\frac{3}{2}+\mu\right)\right]\frac{dF}{dx}+ \nonumber \\
&&\left(-\frac{\mu^2+\mu-l(l+1)}{4x}+\frac{\mu^2+4\mu-l(l+1)+3\omega^2}{4}\right)F=0, \label{equ:32}
\end{eqnarray}
where  $x=R^2$. It corresponds to the well-known hypergeometric equation provided the 
coefficient in the term $F/x$ is zero, i.e, $\mu^2+\mu-l(l+1)=0$. There are two 
solutions for the parameter $\mu$: $\mu_1=l$ and $\mu_2=-(l+1)$. Identifying Eq.~(\ref{equ:32}) with the hypergeometric equation~\citep{Abramovitz72}, two linearly independent solutions write in terms of the hypergeometric 
function ${\cal F}(\alpha, \beta; \gamma, x)$ where the indices $\alpha$, $\beta$ and $\gamma$ are calculated for the two values $\mu_{1,2}$ found above. The general the solution to Eq. (\ref{equ:31}) reads, therefore:
\begin{eqnarray}
&&S(R) = \frac{R^l}{1-R^2}\,C_1\,{\cal F}(\alpha_1,\beta_1;\gamma_1,R^2)\nonumber \\
&&+\frac{R^{-(l+1)}}{1-R^2}\,C_2\,{\cal F}(\alpha_2,\beta_2;\gamma_2,R^2),\label{equ:33}
\end{eqnarray}
where $C_1$ and $C_2$ are two arbitrary constants, and where:
\begin{eqnarray}
&&\alpha_{1,2} = \frac{\mu_{1,2}}{2}+1+\Delta ,\quad \beta_{1,2}  = \frac{\mu_{1,2}}{2}+1-\Delta,\quad \label,\nonumber \\
&&\gamma_{1,2} = \frac{3}{2}+\mu_{1,2}\quad,\label{equ:34}
\end{eqnarray}
with $\Delta = \sqrt{l(l+1)+4-3\omega^2}$. For the given set of indices,  $\cal F$ reduces to the associated Legendre functions~\citep{Abramovitz72}, and we have ${\cal F}(\alpha_1,\beta_1;\gamma_1,R^2) \propto R^{-l-1/2} P(R)/\sqrt{1-R^2}$ and ${\cal F}(\alpha_2,\beta_2;\gamma_2,R^2) \propto R^{l+1/2} Q(R)/\sqrt{1-R^2}$ where $P(R)\equiv P^{l+1/2}_{\Delta-1/2}(\sqrt{1-R^2})$ and $Q(R)\equiv Q^{l+1/2}_{\Delta-1/2}(\sqrt{1-R^2})$. 
As consequence, the solution (\ref{equ:33}) takes the simplified form:
\begin{eqnarray}
&&S(R)=\frac{1}{\sqrt{R}(1-R^2)^{3/2}}\,\left[C_1\, P(R)+C_2\, Q(R)\right].\label{equ:35}
\end{eqnarray}
Next, the components of the perturbed velocity $\delta \h{\vec v} (\h r, \theta, \phi, t)= \delta \h{\vec{V}} (\h r,\theta, \phi)\,\exp\, (\omega \h t/\tau)$ can be expressed in terms of the function $S$:
\begin{eqnarray}
&&\delta \h V_r =\frac{\h a_0}{3\omega\tau}\,\left[-(1-R^2)\,\frac{dS}{dR} + 2R\,S\right]\,Y_{lm},  \nonumber \\
&&\delta \h V_\theta =-\frac{\h a_0}{3\omega\tau R}\,(1-R^2)\,S\,\partial_\theta Y_{lm},  \label{equ:37}\\
&&\delta \h V_\phi =-\frac{\h a_0}{3\omega\tau R\sin\theta}\,(1-R^2)\,S\,\partial_\phi Y_{lm}.   \nonumber
\end{eqnarray}
Coming back to Eq.~(\ref{equ:35}) to get the right $S(R)$, the solution constants $C_1$ and $C_2$ can be found from the boundary conditions~\ref{equ:22}. Dropping the exponential time dependence of the perturbations these boundary conditions provide:
\begin{eqnarray}\label{equ:38a}
&&\left. \h a_0 S(R)= \h a_i\frac{3R}{1-R^2}\right|_{R=R_i}, \nonumber \\
&&\left. R\,(1-R^2)\,\frac{dS}{dR} \right|_{R=R_i}= S\,[2R^2-\omega^2(1-R^2)], \, i=0,1.
\end{eqnarray}
The derivative of $S$ [see Eq.~(\ref{equ:35})] can be found analytically~\citep{Abramovitz72}:
\begin{eqnarray*}
&&\frac{dS}{dR} = \frac{-S(R)}{R(1-R^2)} [(1-R^2)(4-\Delta)-3] \\
&&-\frac{\Delta+l}{(1-R^2)^2R^{3/2}}[C_1 P_1(R)+C_2 Q_1(R)],
\end{eqnarray*}
where  $P_1(R)\equiv P^{l+1/2}_{\Delta-3/2}(\sqrt{1-R^2})$ and 
$Q_1(R)\equiv Q^{l+1/2}_{\Delta-3/2}(\sqrt{1-R^2})$. At the inner boundary, $R=R_0$, equations (\ref{equ:38a}) provide two linear relations between $C_1$ and $C_2$:
\begin{eqnarray*}
&&C_1\,P(R_0)+C_2\,Q(R_0)=3R_0^{3/2}\sqrt{1-R_0^2}, \\
&&C_1\,P_1(R_0)+C_2\,Q_1(R_0)=\frac{3R_0^{3/2}(1-R_0^2)} {\Delta+l}\left(\omega^2+\Delta+\frac{2R_0^2}{1-R_0^2}\right).
\end{eqnarray*}
From these expressions one defines completely the constants $C_1$ and $C_2$ and one gets:
\begin{eqnarray}
&&C_1= \frac{3R_0^{3/2} \sqrt{1-R_0^2}}{{\cal B}}\nonumber \\
&&\left[Q(R_0) \frac{ \sqrt{1-R_0^2}} {\Delta+ l} \left(\omega^2 +\Delta+ \frac{2R_0^2-1}{1-R_0^2}\right) - Q_1 (R_0)\right], \label{equ:41} \\
&&C_2= \frac{3R_0^{3/2} \sqrt{1-R_0^2}}{{\cal B}}\nonumber \\
&&\left[P(R_0) \frac{ \sqrt{1-R_0^2}} {\Delta+ l} \left(\omega^2 +\Delta+ \frac{2R_0^2-1}{1-R_0^2}\right) - P_1 (R_0)\right], \label{equ:42}
\end{eqnarray}
where ${\cal B }= P_1(R_0) Q(R_0) -P(R_0) Q_1(R_0)$. From Ref.~\citep{Abramovitz72}, one can show that $\cal B$ is a constant equal to $-\Gamma (\Delta+l)/\Gamma (\Delta -l)$ where $\Gamma$ is the classical Euler gamma function. 

With $C_1$ and $C_2$ given by Eqs.~(\ref{equ:41}) and (\ref{equ:42}), the solution (\ref{equ:35}) is completely defined. It depends on the mode number $l$, on the position of the inner surface $R_0$ and on the free parameter $\omega$. The latter should be defined from the remaining boundary conditions at the outer interface $R=R_1$. The perturbations of the motion both interfaces ($R=R_0$ and $R=R_1=1$) cannot be taken initially  independently, and cannot further evolve independently. Indeed, from first equation (\ref{equ:38a}) at the outer surface $R \equiv R_1=1$ one finds:
\begin{eqnarray}\label{equ:43}
&&\h a_1= \frac13\, \h a_0 \lim_{R\to 1} [S(R)\,(1-R^2)]. 
\end{eqnarray}
Notice that this relation does not forbid the calculation of the divergence of $S$ near the outer boundary, because the density and the sound speed of the unperturbed flow  go to zero there. Since in Eq.~(\ref{equ:35}), $P$ and $Q$ are rational functions of $y =\sqrt{1-R^2}$, one can expand $S$ in Taylor series for $y\to 0$ and  Eq.~(\ref{equ:43}) becomes:
\begin{eqnarray}\label{equ:44}
&&3\,\frac{\h a_1}{\h a_0}= \nonumber \\
&&\lim_{y\to 0} \left[\frac{C_1 P(1) +  C_2 Q(1)}y + C_1 \left. \frac{dP}{dy}\right |_{y=0} +  C_2 \left. \frac{dQ}{dy}\right |_{y=0} \right]. 
\end{eqnarray}
The first term in the right hand side of (\ref{equ:44}) diverges if the numerator is non-zero. 
Therefore, the only way to satisfy this boundary condition and to obtain a finite displacement at the outer surface is setting the numerator to zero. This condition $C_1 P(1) +  C_2 Q(1)=0$ corresponds to the dispersion relation for the shell instability we are looking for. Using the explicit formulae for $C_1$ and $C_2$, we obtain:
\begin{eqnarray}
&&\omega^2=-\Delta -\frac{2R_0^2 }{1-R_0^2} \nonumber \\
&&-\frac{\Delta +l }{\sqrt{1-R_0^2}} \,\frac{P(1) Q_1(R_0) - Q(1) P_1(R_0)}{P(R_0) Q(1) - P(1) Q(R_0)}.\label{equ:45}
\end{eqnarray}
It will be shown later in Sec.~\ref{Sec:6} that provided (\ref{equ:45}) is satisfied, the second boundary condition at the outer edge is also immediately satisfied.

Equation (\ref{equ:45}) provides a relation between the growth rate of the instability, $\omega$, and the mode number $l$. It is independent of the azimuthal number $m$, because the unperturbed system is spherically-symmetric and it will be shown later that, in addition, it does not depend on the shell inner radius $R_0$.

Moreover, from Eq.~(\ref{equ:44}) one obtains also a relation between the displacements at the shell surfaces:
\begin{eqnarray}\label{equ:46}
&&3\,\frac{\h a_1}{\h a_0}= 2^{l+1} \frac{\Gamma[(\Delta+l)/2+1]}{\Gamma[(\Delta-l)/2]}\nonumber \\
&&\left( C_1 \sqrt{\frac2\pi} \sin\frac{\pi(\Delta+l) }2 +  C_2 \sqrt{\frac\pi2} \cos\frac{\pi(\Delta+l) }2\right). 
\end{eqnarray}
This will be used later on to analyze the temporal evolution of the instability.

\section{Numerical solution of the dispersion relation}\label{Sec:5}

\begin{figure}
\centerline{\includegraphics[width=8cm]{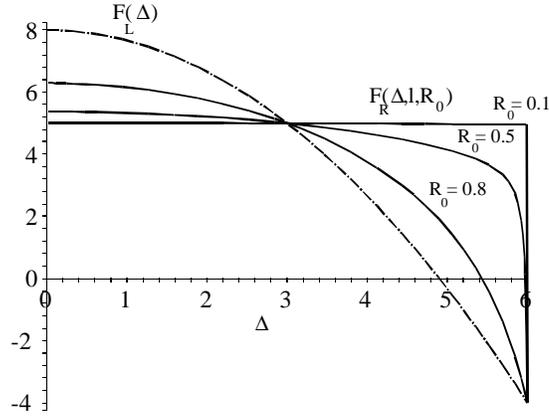}}
\caption{Plot of the two functions  $F_L(\Delta)$ and
$F_R(\Delta,l,R_0)$ representing the left and the right hand sides
of Eq. (\ref{equ:45}). The intersection of these curves provides
the value of $\Delta$, corresponding to the solution of the
dispersion equation (\ref{equ:45}). Both curves depend on
the mode number $l$ and $F_R$ depends also on the shell thickness
$R_0$ (curves are drawn for $R_0=0.1, 0.5$ and $0.8$). However, we see that the intersection point does not move. In this example $l=4$, 
the value of $\omega^2$ is $5$ and $-4$, i.e., $l+1$ and $-l$ and we see that the intersection point is located at $\Delta=3$.\label{fig5}}
\end{figure}

The dispersion equation (\ref{equ:45}) looks rather complicated and we first consider its numerical solution. For this purpose, we represent it as a relation between two functions, $F_L(\Delta)=F_R(\Delta)$, depending on the parameter $\Delta$. Here, $F_L(\Delta)=l(l+1)+4-\Delta^2 = \omega^2$ is the left hand side of Eq.~(\ref{equ:45}) and $F_R(\Delta)$, its right hand side. In Fig.~\ref{fig5}, $F_L$ and $F_R$ are plotted as a function of the variable $\Delta$ and we search for the intersection points of these two functions for given parameters $l$ and $R_0$. In order to have an unstable solution ($\omega^2$ should be real and positive) the intersection must occur in the upper right part of the plane $(F_{R,L};\Delta) : 0 <\Delta<\sqrt{l(l+1)+4}=\Delta_m$ and  $0<F_R<l(l+1)+4$.  This defines the upper limit for the growth rate for a given $l$: $\omega^2 < [l(l+1)+4]/3$.  To have a stable solution, the intersection must occur in the lower right part of the plane $(F_{R,L};\Delta):0 < \Delta < \sqrt{l(l+1)+4}= \Delta_m$ and $F_R <0$. These  restrictions are useful to constrain the numerical solution of the dispersion equation.

An example of curves $F_R, F_L$ is shown in  Fig.~\ref{fig5}. Both curves are surprisingly simple. The left function $F_L$ is just a parabola which has a maximum at $\Delta = 0$ and becomes negative for $\Delta>\Delta_m$. The right hand side has also a maximum at $\Delta = 0$ and it decreases with $\Delta$. There are two intersection points: one corresponds to $\omega^2>0$ (unstable mode) and another one to $\omega^2<0$ (oscillating mode at $\Delta=6$ for the present mode). The numerical solution indicates that both intersection points are independent of the shell thickness $R_0$. Especially, the point corresponding to unstable solutions is always located at $\Delta=3$ for chosen value $l=4$. It will be shown below that this is a consequence of the nature of the perturbed flow.

\section{Behavior of the perturbed flow}\label{Sec:6}

Until now, we have considered a general solution for the perturbed flow, without any assumption about its nature. Especially, we have been working with equations including compressible effects. However, this rather simple solution for the instability growth may suggest that special properties should probably arise. In this section we are going to study the properties of the perturbed flow.

Since $\partial_{\theta}(\sin \theta \partial_{\theta} Y_{lm})/\sin \theta+(\partial^2_{\phi} Y_{lm})/\sin^2 \theta = -l(l+1) Y_{lm}$, and after the injection of the second derivative of $S$, $d^2S/dR^2$ obtained from Eq.~(\ref{equ:31}),  the divergence of the perturbed velocity $\de \h{\vect{V}}$ [given by Eqs. (\ref{equ:37})] reads,
\begin{eqnarray}
&&\diver  \de \h{\vect{V}} = \frac{\h a_0\,\omega}{\h r_1 \tau}
\left\{-S - \frac{R}{\omega^2} \left(\frac{d S}{dR}-\frac{2RS}{1-R^2}\right)\right\} Y_{lm}\, .\label{equ:48}
\end{eqnarray}
It is worth to note that from the second boundary condition (\ref{equ:38a}) the velocity divergence is zero at the boundaries. Now, we make a stronger assumption. Let us assume that the flow is incompressible everywhere, i.e., $\diver \de \h{\vect{V}} =0$ for any value of the radius $R$. Then, from (\ref{equ:48}) we can express the first derivative of $S$, $dS/dR$, in terms of $S$ ans $R$ and we can compute $d^2S/dR^2$, as well. Plugging these derivatives in the perturbation equation (\ref{equ:31}), we find after some simplifications that the coefficient in front of $S$ does not depend on $R$ and the differential equation reduces to a simple algebraic relation between $\omega$ and $l$ :
\begin{eqnarray}
&& \omega^4-\omega^2-l(l+1)=0 \label{equ:49}. 
\end{eqnarray}
This relation is a consistency condition between the eigenmode equation
(\ref{equ:31}) [it is automatically satisfied] and our additional constraint coming from (\ref{equ:48}), i.e., requiring that $\de {\h{\vect{V}}}=0$ everywhere inside the shell. It is also consistent with the boundary conditions (\ref{equ:38a}).
Therefore Eq.~(\ref{equ:49}) can be considered as the dispersion relation. We have to admit that up to now we have not been able to find a solution to Eq.~(\ref{equ:31}), satisfying (\ref{equ:38a}), that differs from the one providing a zero divergence of the perturbed velocity. In other words the solution that is ``constrained'' by the boundary conditions (\ref{equ:38a}) describes an incompressible RT perturbation for which the dispersion relation is given by (\ref{equ:49}).

The solutions of this quartic polynomial equation read easily :
\begin{eqnarray}\label{equ:49a}
&&\omega_{1,2}=\pm\sqrt{l+1}, \quad   \omega_{3,4}=\pm i \sqrt{l}.
\end{eqnarray}
Moreover, it should be emphasized that these solutions perfectly agree with the solutions of the general dispersion equation (\ref{equ:45}) found numerically in the previous section. 
As a matter of fact, for $l=4$ we have found numerically $\omega^2=5$ and $\omega^2=-4$ leading to $\omega=\pm \sqrt{5}$ and $\omega=\pm 2i$, respectively. These values are those exactly obtained from (\ref{equ:49a}) by replacing $l$ with the specific value $l=4$.

\begin{figure}
\centerline{\includegraphics[width=8cm]{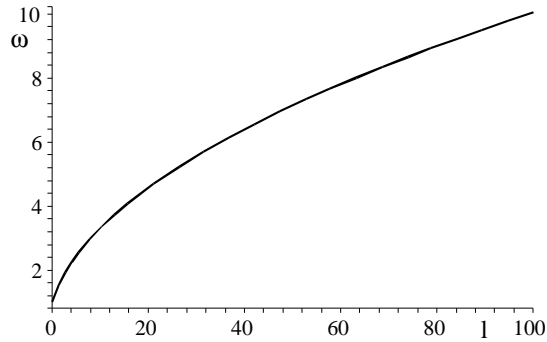}}
\caption{Dependence of the dimensionless growth rate $\omega$ on
the mode number $l$, $\omega=\sqrt{l+1}$ [from Eq.~(\ref{equ:49a})].\label{fig6}}
\end{figure}

As consequence, the solution we have found assuming a velocity profile with a divergence equal to zero does not look anymore as a ``trick'' to satisfy the whole set of constraints. This result is the proof that the RT unstable flow is indeed incompressible and the growth rate is unexpectively independent on the shell thickness.  This result is quite new and is shown here in this paper for the first time, although some people studied before in details thin and thick shells (see for instance the monography by Kull~\citep{Kull91}. All the modes are unstable for any value of the mode number $l$ and the most unstable growth rate, $\omega(l)$, among $\omega_1(l)$ and $\omega_2(l)$, is shown in Fig.~\ref{fig6}. 

In addition, one can easily check that the flow is irrotational. This property arises since the unperturbed flow is radial. The vorticity of the perturbed flow in the laboratory frame (\ref{equ:5}) is given by: $\rot \de \vec{v} = \rot \de\h{\vec{v}} /C(t)$ since $A(t)=C(t)$. As $\rot\de\h{\vec{v}}=0$ from Eqs.~(\ref{equ:37}), the perturbed flow is therefore also irrotational in the ``physical'' space.

\section{Evolution of the shell perturbations}\label{Sec:7}

The effect of the instability can be further analyzed by observing the deformation $\h\xi$ of the shell. It can be calculated from the velocity field in the rescaled frame by using the relation $d \h{\vec \xi} / d \h t = \de \h {\vec v}$ which provides $\h{\vec{\xi}}(\h r,\theta,\phi,\h t)= (\tau/\omega)\de \h {\vec v} = (\tau/\omega)\exp(\omega \h t/\tau)\de \h {\vec V}(\h r,\theta,\phi)$, where $\de \h {\vec V}$ is given by (\ref{equ:37}). Since the dispersion equation has four roots (\ref{equ:49a}), the general solution for the displacement is a linear superposition of four modes. In particular, for the radial displacement, one has :
\begin{eqnarray}
&&\h \xi_r(R,\theta,\phi,\h t)=\h a_0\,\sum_{i=1}^{4} \Lambda_i \h \xi_i(R)\,{\rm e}^{\omega_i \h t/\tau}\, Y_{lm}(\theta,\phi) , \label{equ:58} 
\end{eqnarray} 
where
\begin{eqnarray*}
&& \h \xi_{i}(R)=\frac{1}{3\omega_i^2} \left[-(1-R^2)\,\frac{dS_i}{dR} + 2R\,S_i\right], 
\end{eqnarray*}
are the displacement eigenmodes and $\Lambda_i$ are the mode amplitudes
($\Lambda_{1,2}$ are real, $\Lambda_3=\Lambda_4^*$). Moreover, as the function $S$ depends on $\omega^2$ then $\h \xi_1=\h \xi_2$ and $\h \xi_3=\h \xi_4$. We have $\h \xi_i(R_0)=1$ also. This equality comes from the normalization of $\h \xi_i(R)$ with the inner initial displacement $\h a_0$ (see Section~\ref{Sec:4}). Then $\h \xi_i(1)$ is given by Eq.~(\ref{equ:46}). 

At this stage, we can calculate the deformation of the shell in the initial space $(R,\theta,\phi,t)$. From the scaling relation for the distances [first equation of Eqs.~(\ref{equ:3})] and from the relation (\ref{equ:13}) between $t$ and $\h t$, the radial displacement $\xi_r(R,\theta,\phi,t)$ writes :
\begin{eqnarray}
&&\xi_r(R,\theta,\phi, t)= a_0\,C(t)  Y_{lm}(\theta,\phi) \sum_{i=1}^{4} \Lambda_i \h \xi_i(R) {\rm e}^{\omega_i g(t)} , \label{equ:58b} 
\end{eqnarray}
where $a_0=C(0)\h a_0= \h a_0$. 
The radial velocity perturbation is provided by Eq.(\ref{equ:5}) with $A(t)=C(t)$ since $\gamma=5/3$ : 
\begin{eqnarray}\label{equ:59}
&&\de \vec v = \frac{\de \h{\vec{v}}} {C(t)}= \frac{\de \h{\vec{V}}} {C(t)} {\rm e}^{\omega g(t)},  
\end{eqnarray}
where $\de \h {\vec{V}}$ is given by Eqs.~(\ref{equ:37}).
The derivation of (\ref{equ:59}) requires further details. The velocities $\vec v(\vec r,t)$ and $\h{\vec v}(\h{\vec r},\h t)$ in the $(\vec r, t)$ and $(\h{\vec r},\h t)$ spaces, respectively, are connected through Eq.~(\ref{equ:5}). In $(\h{\vec r},\h t)$, we have $\h{\vec v}(\h{\vec r},\h t)=\de \h{\vec v}$, while the corresponding equation is $\vec{v}(\vec r, t)=\de \vec v+\vec v_0 (\vec r, t)$ in $(\vec r, t)$ where $\de \vec v$ is the velocity perturbation and where $\vec v_0(\vec r, t)$ is the background velocity given by $\vec v_0 (\vec r,t)=\dot C(t) \vec r / C(t)$. Introducing these expression in (\ref{equ:5}), one obtains (\ref{equ:59}) immediately.
In addition, we are going to introduce the quantity $\de \vec V(\vec r,t)$ related to $\de \vec v$ by $\de \vec v = \de \vec V \, \exp[\omega g(t)]$ (this relation is similar to the relation between $\de \h{\vec v}$ and $\de \h{\vec V}$).

It is now possible to derive the constants $\Lambda_i\,(i=1,4)$ from the initial velocity perturbation profile in the physical space. 
In the hat space, the radial velocity $\de \h v_r$ is given by $\de \h{v_r}= d \h \xi_r/d\h t$. On the other hand since we have $\de V_r (R,0)= \de \h V_r(R,0)$ at $t=\h t=0$ and for $\theta=\phi=0$, the $\Lambda_i$'s can be calculated from the boundary conditions taken as $\xi_r(R_0,0)=a_0\,D_0$, $\xi_r(1,0)=a_0\,D_1$, $\de V_r(R_0,0)= a_0\,V_0$, $\de V_r(1,0)= a_0\,V_1$ in the physical space where $D_0,D_1,V_0$ and $V_1$ are four arbitrary parameters. Solving the corresponding system of equations, we find :
\begin{eqnarray} 
&&\Lambda_{1,2}=\frac{1}{2(b_1-b_2)}\left[D_1-D_0 b_2 \pm \frac{\tau}{\sqrt{l+1}}(V_1-V_0 b_2)\right], \label{equ:64}\\
&&\Lambda_{3,4}=\frac{1}{2(b_1-b_2)}\left[D_0 b_1-D_1 \mp i \frac{\tau}{\sqrt{l}}(V_1-V_0 b_1)\right], \label{equ:65}
\end{eqnarray}
where $b_1=\h \xi_{1,2}(1)$ and $b_2=\h \xi_{3,4}(1)$ and the
eigenmodes are normalized by the condition $\h \xi_{i}(R_0)=1$ $(i=1,4)$. In
the following we study the evolution of several specific perturbations. 
 
\begin{figure}
\centerline{\includegraphics[width=8cm]{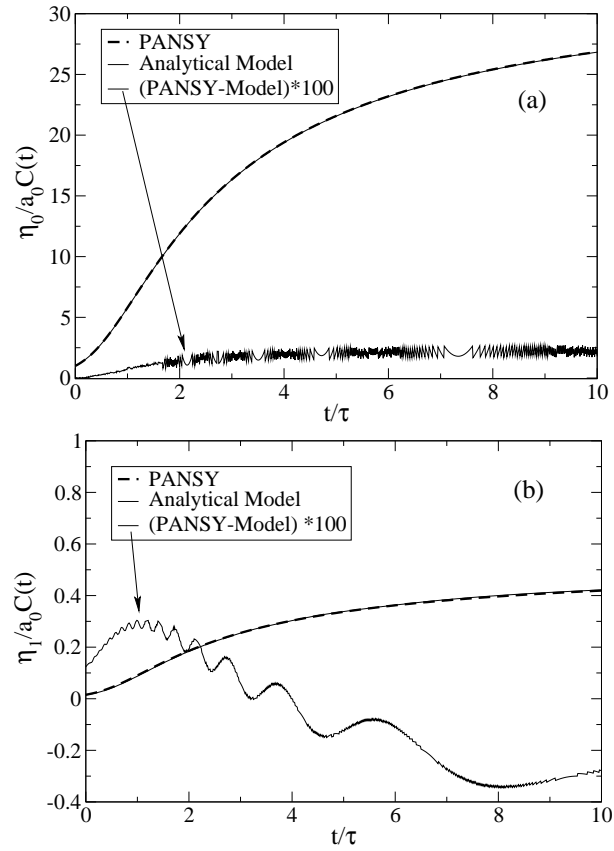}}
\caption{Comparison between the time evolution of the inner (a) and outer (b) displacements of the shell for the more unstable mode: $\Lambda_1=1$ and $\Lambda_{2,3,4}=0$ in the case $l=4$ with a inner radius $R_0=0.5$ and $\beta=0$. The displacement is divided by the scale function $C(t)$  to exhibit clearly the perturbation growth due to the RTI. These plot show a good agreement between the analytical solution (solid line) and the simulation (dashed curve) with the perturbation code PANSY. The difference between the two results is less than one percent.\label{fig7}}
\end{figure}

First, we study the most unstable mode, $\omega_1=\sqrt{l+1}$ by setting $\Lambda_1=1$ and $\Lambda_{2,3,4}=0$. The solid lines in Fig.~\ref{fig7} show the time evolution of the radial shell normalized displacement at the inner (a) and outer (b) interface, $\eta_{0,1}/a_0 C(t)$, in the laboratory frame. It depends on two parameters: $\tau$ - the characteristic expansion time and $\beta$ - the initial shell velocity. The exponential law is valid only for $t\ll\tau$ (acceleration phase of the shell), indeed, $\exp[\omega_1 g(t)] \rightarrow \exp(\omega_1 t/\tau)$, then $\eta_{0,1}/a_0=\exp(t/\tau)$, with $C(t)\rightarrow 1$. For $t\gg\tau$, $\exp g(t) \rightarrow \exp[\omega_1(\pi /2-\arctan{\beta})]$ it following that $\eta_{0,1}/a_0(t/\tau)=\exp[\omega_1(\pi /2-\arctan{\beta})]$ with $C(t)\rightarrow t/\tau$. The shell is in a ballistic motion and the perturbation increases linearly, i.e., $\eta_{0,1}\propto t/\tau$ - the function $g(t)$ is bounded and $C(t)$ is a linear function of time. In the figure~\ref{fig7} the displacement is divided by the scale function $C(t)$ to exhibit clearly the perturbation growth due to the RTI. It is interesting to note that in the case of zero initial velocity the amplification factor e$^{\omega \,\pi/2}$ does not depend on the expansion time. Moreover, the amplification is smaller if the shell is always in expansion. In opposition, the amplification can be stronger for initially converging shell ($\beta < 0$ but $\mid \beta \mid \gg 1$).

The numerical model/code employed here is called Pansy~\citep{Pansy}. It computes the time development of three-dimensional modes of coupled hydrodynamic, thermodynamic, and transport phenomena, including heat flow, viscosity, fully linearized about zeroth order spherically or cylindrically symmetric compressible flows. The zeroth order solutions are calculated on a typical one-dimensional lagrangian grid and have the form $f^j(t)$, where the $f$' s are all of the necessary hydrodynamic variables, and other variables including zone radius, and where $j$ is the radial zone index. First order quantities, of the form $f_{l,m}^j(t) Y_{l,m}(\theta) \exp{\left(im\phi\right)}$ for spherical geometry, are calculated with difference equations which are linearly perturbed forms of the former discretized zeroth order equations, rather than discretizations of the linearly perturbed continuous zeroth order equations. This relatively conservative/Hamiltonian differencing approach produces considerably improved treatment of phenomena requiring high resolution, especially artificial viscosity for shocks in contrast with earlier form of Pansy which required higher resolution for the same accuracy~\citep{Henderson74}.

The analytical solution for the inner and outer shell displacement
of the unstable mode has been compared with the simulations
performed with the perturbation code Pansy.
The agreement between the analytical solution and the simulations 
is very good. The difference is less than one percent. It is shown in
Fig.~\ref{fig7} that the inner interface growth is more rapid because
this surface is RT unstable, while the outer interface is in
contact with the vacuum and does not show a substantial growth. It
looks stable. The deviation between the theory and simulations at the
outer surface is larger than at the inner interface because it is more 
difficult for the code to handle the contact with vacuum.

From Eq.~(\ref{equ:13}), the amplification factor $\exp[\omega g(t)]$ for $t \rightarrow +\infty$ is given by $\exp[\sqrt{l+1}(\pi/2-\arctan \beta)]$. For $\beta=0$, we recover the special case studied by  Bernstein and Book~\citep{Bernstein78} and the value they found is very close to the one we have. For example, for $l=40$, Book and Bernstein calculated numerically the amplification $\simeq10^4$, while our formula provides  $\exp(\sqrt{l+1} \pi/2)=2\times 10^4$.
 
The asymptotic amplification depends strongly on the initial velocity defined by the parameter $\beta$ [see Eq.~(\ref{equ:16})]. For example, if $\beta=1$, the shell experiences a monotonic expansion phase, and the amplification decreases to $10^3$. Therefore, the shell is less fragile. However, for $\beta =-1$ the shell collapse initially and the amplification would be $10^8$. As a result, the shell becomes much more fragile. We conclude therefore that most dangerous stage of the shell evolution is the stagnation phase, which has been analyzed by Hattori et al.~\citep{Hattori86}.

The purely growing eigenmode that we have studied above
corresponds to correlated perturbations between both surfaces (inner
and outer) which seems to be not too realistic. It would be more
appropriate to consider either independent initial perturbations of
the inner and outer surfaces or the perturbation of the inner
surface only. According to (\ref{equ:64}) and (\ref{equ:65}), from
four incompressible modes one can construct any kind of initial
perturbation. Let us consider a case where the inner and outer
interface displacements are opposite, i.e., $D_0=1$, $D_1=-1$, $V_0
= V_1 = 0$. Hereafter we call this type of perturbation the
``sausage''. This ``sausage'' could be produced in the outer shell of the star by
convection phenomena arising just before its explosion.
 
\begin{figure}
\centerline{\includegraphics[width=11cm]{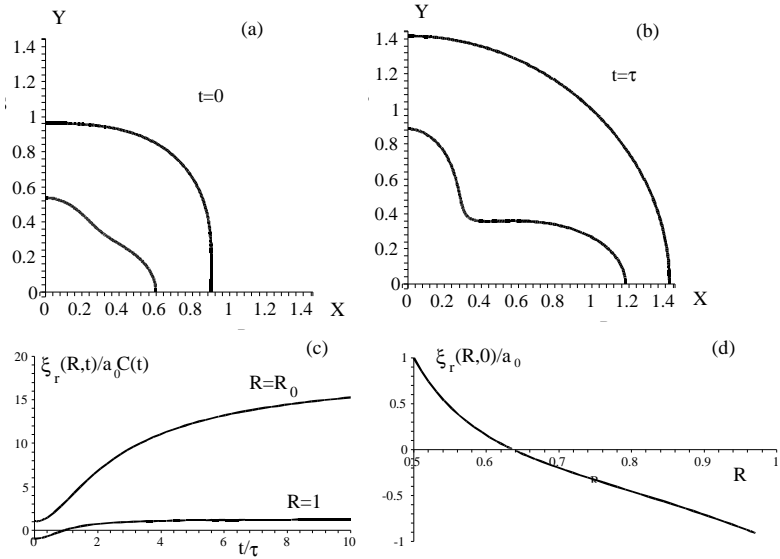}}
\caption{Plot of the initial ``sausage'' shell perturbation (a) at
$t=0$ and (b) at $t=\tau$. Panel (c) shows the time evolution of the
inner and outer shell displacements and (d) gives the initial
displacement within the the shell. The mode number is $l=4$ and the
shell thickness is $R_0=0.5$.\label{fig8}}
\end{figure}

From Eqs.~(\ref{equ:64}) and (\ref{equ:65}) one finds:
$\Lambda_1=\Lambda_2=(-1-b_2)/[2(b_1-b_2)]$ and 
$\Lambda_3=\Lambda_4=(b_1+1)/[2(b_1-b_2)]$.
The radial profile of the initial displacement within the shell is
shown in Fig.~(\ref{fig8}d). The time evolution of such a shell
perturbation in the laboratory frame is shown in Fig.~(\ref{fig8}c)
for the mode $l=4$. We have normalized the displacement $\xi_r$ by the
scale function $a_0\,C(t)$ just to exhibit the amplification due to the
RTI and suppress the shell thickness expansion. The
growing perturbation at the inner surface of the shell due to the
RTI is clearly shown, while the outer surface becomes spherical [see 
Fig.~(\ref{fig8}b)] because this interface is stable. The perturbation grows for time period satisfying $t<\tau$ when the shell is accelerated and one can see that the displacement is larger along the polar axis $\theta=0$ [$X$ axis on Fig.~(\ref{fig8}b)].

Additional types of shell perturbations are interesting to study. Three of then exist : (i) the ``kink'' configuration for which the inner and outer initial perturbations are the same, $D_0=D_1=1$, $V_0 = V_1 = 0$; (ii) the ``inner'' shell perturbation for which only the inner interface is initially perturbed, $D_0=1$, $D_1=0$, $V_0 = V_1 = 0$; (iii) the ``velocity'' alteration for which the positions of the inner and outer surfaces are initially unperturbed but just the inner velocity is perturbed, $D_0=D_1=0$, $V_0=1$, $V_1=0$. This configuration describes, for instance, a wind pulsar flux fluctuation.

The relative weight of the unstable mode in each of perturbation is given by the constant $\Lambda_1$. Hence, this parameter can be used to compare the configurations. For the ``sausage" case with $R_0=0.5$ and $l=4$ shown in Fig.~\ref{fig8}, $\Lambda_1=0.56$. For the three other shell perturbations we find : $\Lambda_1=0.43$ (``kink"), $\Lambda_1=0.50$ (``inner''), and $\Lambda_1=0.22$ for the ``velocity'' case. Therefore, all the three displacement perturbations configurations, ``sausage'', "kink" and "inner" are equally unstable. However, the amplification coefficient for these three modes is approximately twice smaller than for the growing single mode ($\Lambda_1=1$ and $\Lambda_j=0, j=2,4$). These modes can probably be excited more naturally from the supernova progenitor heterogeneities. Finally, the "velocity" perturbation seems to be the less unstable. Its amplification coefficient is more than four time smaller.

However, the respective importance of different perturbations depend on the shell thickness and the mode number. For example, for $R_0=0.9$ and the same mode number $l=4$ we find: $\Lambda_1=1.41$ for the ``sausage'' mode, $\Lambda_1=0.22$ (``kink''), $\Lambda_1=0.80$ (``inner''), and $\Lambda_1=0.36$ (``velocity''). In this case, the ``sausage'' mode is the most dangerous and it grows even stronger than the pure mode.
  
By knowing the mode amplification coefficient, one can conclude about the shell fragility. In other words, we are able to derive a supernova shell fragmentation criterion due to the RTI. Assuming that the disruption of the shell occurs when the perturbation amplitude becomes comparable to the shell thickness, this criterion defines the critical amplitude, $a_{cr}$, of the initial perturbation for the shell disruption. We have immediately this criterion in the hat space, with Eq.~(\ref{equ:58}) for $\theta=\phi=0$, the displacement of the inner interface on the X axis for the more unstable mode becomes : $\h \xi_r(R_0,0,0,t)= a_{cr} \Lambda_1 \exp [\omega_1 g(t)]= (\h r_1-\h r_0)=\h r_1\,(1-R_0)$. We can rewrite this criterion as : 
\begin{eqnarray}\label{equ:70}
&&   \frac{a_{cr}}{\h r_1} \approx \frac{1-R_0}{\Lambda_1}\,{\rm e}^{-g(t)\sqrt{l+1}}, \, t \rightarrow +\infty.
\end{eqnarray}
This critical amplitude depends on the mode number $l$, the shell thickness, and the type of perturbation. The formula is valid for low modes, $l < l_M = 2\pi R_0/(1-R_0)$, with a breaking occurring in the linear regime. The higher modes, $l > l_M$, seem to be less dangerous. Although their amplification coefficient can be large, the corresponding perturbation enters in non-linear regime and grows less rapidly. In contrast, the lower modes, $l<l_M$, are growing slowly and they lag behind the mode $l_M$. Therefore, the linear theory predicts that the most dangerous perturbation corresponds to the mode $l_M$ and the critical initial amplitude is given by Eq.~(\ref{equ:70}) where $l$ is replaced with $l_M$. For example, for $R_0=0.65$ and $\beta =0$  the most dangerous mode is $l_M \sim 12$ with the ``sausage'' or ``inner'' type of perturbation ($\Lambda_1 \approx 0.5$). Then we obtain $a_{cr}/\h r_1= 2.4 \times 10^{-3}$. The ``velocity'' mode  is much less dangerous since it corresponds to $\Lambda_1=0.14$.

\begin{figure}
\centerline{\includegraphics[width=8cm]{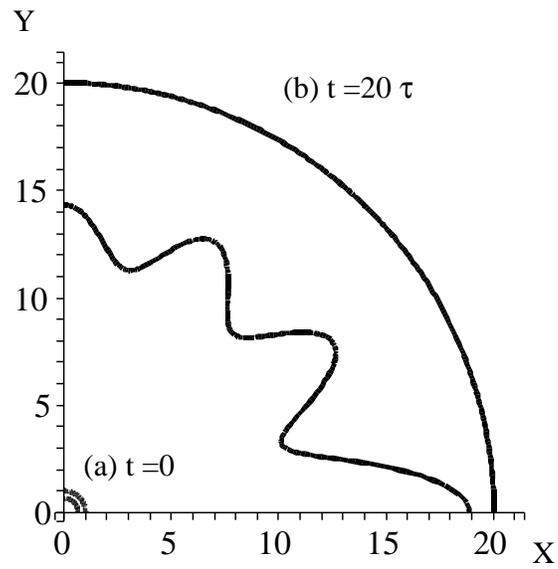}}
\caption{Plot of the initial inner shell perturbation at $t=0$
(a) and $t=20\tau$ (b). The mode number is $l=12$ and the shell
thickness is $R_0=0.65$. The relative initial inner shell perturbation is
0.24\%.\label{fig9}}
\end{figure}

As result, a perturbation amplitude as small as $0.24\%$ of the inner radius would be large enough to disrupt the shell. Figure~{\ref{fig9}} shows the initial shell with the critical perturbation $a_{cr}$ for a ``inner'' configuration (panel a) and the shell close to disruption at time $t=20\tau$ (panel b). The theory predicts that the shell will break down near the pole (X-axis).

\section{Discussion and conclusion}\label{Sec:8}

In the first part of this paper, a non-stationary spherical flow
describing the expansion of SN ejecta has been derived by applying
a rescaling method~\citep{Bouquet85} to the Euler
equations for a polytropic gas. This is a non-stationary solution
that describes the motion of a shell with initial finite
thickness $L_0$. The shell dynamics is described by two
parameters: the initial velocity $\beta$ and the expansion time
$\tau$. By changing the sign of $\beta$ one may study implosions
and explosions.

This solution is relevant for the description of the plerion
evolution -- type II SN remnant driven by a central pulsar
pressure. The temporal behavior of the pressure in our model is in agreement with the classical spin-down power law of the pulsar luminosity~\citep{Blondin01}. The parameter $\beta$ takes into account the initial kinetic
energy of the shell, which is released during the explosion, and
the time $\tau$ describes the life time of the pulsar at the center of the ejecta.

The transformation from the laboratory frame to the co-moving reference frame allowed us to perform an analytical study of the 3D linear stability of this time-dependant radial flow. The linearized hydrodynamic equations for a mono-atomic gas ($\gamma=5/3$) have been solved using a decomposition of the perturbation in spherical harmonics. The dispersion relation defines the growth rate $\omega$ of any mode $l$ in the co-moving frame, which corresponds to a finite amplification in the laboratory frame. The growth rate is independent of the shell thickness $R_0$ and the azimuthal mode number $m$. Although no assumptions have been made concerning the perturbed flow, we found that the unstable perturbation is incompressible and irrotational.  An analytical expressions for the growth rate and the mode structure were confirmed by comparison the analytical theory with numerical simulations performed with the perturbation code PANSY~\cite{Pansy}. The sign and the magnitude of the initial velocity play an important role in the RTI development. High initial velocities stabilize the shell. In opposition, for the case of initially collapsing shell, $\beta < 0$, the perturbation grows to much higher amplitudes.

By using a linear superposition of stable and unstable modes we studied the dependence of the amplification coefficient on the initial decomposition. It was found that the deformation for a thick shell of the inner interface is most dangerous and corresponds to a amplification coefficient twice smaller than for the unstable single mode. While for a thin shell the ``sausage'' configuration seems to be the more dangerous mode. For a given shell thickness $R_0$, we defined the mode $l_M$ which is the most dangerous for the shell disruption. A criterion is derived that defines the critical initial inner surface deformation that produces a fragmentation of the shell.

A filamentary structure in the Crab nebula expansion has been observed by Hester,~\citep{Hester96} Sankrit et al.~\citep{Sankrit98}. The structure of radial filament velocities indicates that they have a common origin - the SN spherical shell. Such a filamentary structure has been reproduced from numerical simulations by Jun~\citep{Jun98}, who shows that the RTI driven by the acceleration of the thin shell provides the main mechanism for the shell disruption. Our analytical model is in a good agreement with these observations and numerical simulations. By considering a shell with the aspect ratio 10 ($R_0=0.9$), the same as in the paper~\citep{Jun98}, we deduce that most dangerous mode is $l_M=60$. This agrees with the fastest growing mode observed in Jun's simulations. Moreover, the amplitude of initial perturbations ($\sim 1\%$) used in~\citep{Jun98} follows from Eq.~(\ref{equ:70}) for the realistic initial parameters: the velocity of shell is 500 km/s,  the initial radius 0.2 pc, and the explosion time $\tau=500$ yrs (this implies $\beta=1.3$). This demonstration suggests that our analytical model is relevant to study the stability analysis of various SN remnants.

Finally, this work can be applied to ICF or, more generally, to study of RTI in laser target design.

In this respect, this study is directly useful to Laboratory Astrophysics issues and it can be used to design appropriate laser target to examine the RTI problem in SNR~\cite{Ribeyre_Thesis}.

\begin{acknowledgements}
One of us (J.S.) acknowledges the hospitality of the CEA. This research was partially supported by the CICYT of Spain (FTN 2000-20048-C0301) and by the ``Secretar\'ia de Estado de Educaci\'on y Universidades de Espana'' (Programa de Sab\'aticos). 
\end{acknowledgements}

\newpage

\clearpage
\end{document}